\documentclass{vgtc}                          %
\graphicspath{{figures/}{pictures/}{images/}{./}} %
\usepackage{wrapfig, lipsum}
\usepackage{times}                     %
\usepackage{tabu}                      %
\usepackage{booktabs}                  %
\usepackage{lipsum}                    %
\usepackage{mwe}                       %
\usepackage{enumerate, enumitem}
\usepackage{mathptmx}                  %

\vgtccategory{Research}

\vgtcpapertype{Empirical Study}

\vgtcinsertpkg

\title{What Do We Mean When We Talk About \Igx?}
\author{Xiaoyu Liu\thanks{\{dasqxyl, fy, leozcliu\}@umd.edu}\hspace*{-10pt}%
\and Vishnu Sreekanth\thanks{\{vishnus, vpatel08, smuzumda, dlopez22\}@terpmail.umd.edu\vspace*{-40pt}}\hspace*{-10pt}%
\and Vraj Patel\footnotemark[2]\hspace*{-10pt}%
\and Shalin Muzumdar\footnotemark[2]\hspace*{-10pt}%
\and Diego Lopez\footnotemark[2]\hspace*{-10pt}%
\and Fumeng Yang\footnotemark[1]\hspace*{-10pt}%
\and Zhicheng Liu\footnotemark[1]}
\affiliation{\vspace{-6pt}
\scriptsize University of Maryland, College Park
\vspace{-15pt}}

\input{sec/import}

\newcommand{\eg}{\mbox{e.g.,}\xspace\@}

\newcommand{\hci}[0]{\textsc{hci}\@\xspace}
\newcommand{\sm}[0]{supplementary materials\@\xspace}

\makeatletter

\renewcommand\paragraph[1]{%
  \@startsection{paragraph}{4}{\z@}
  {.2em \@plus .2em \@minus.0em}%
  {-.5em}%
  {\normalfont\bfseries}%
  {#1}%
}
\makeatother

\newcommand{\anotherpoint}{\ptr~}

\usepackage[most]{tcolorbox}

\newtcbox{\tagin}[1]{
    on line,
    arc=1.5pt,
    colback=#1,
    colframe=#1,
    before upper={\rule[-1pt]{0pt}{7.5pt}},
    boxrule=.5pt,
    boxsep=0pt,
    left=1pt,
    right=1pt,
    top=-.5pt,
    bottom=.5pt,
    colupper=white
}

\newcommand{\colortag}[2]{\raisebox{0.5pt}{\tagin{#1}{{\textsf{\fontsize{5pt}{5pt}\selectfont{#2}}}}}}

\abstract{%

{There has been limited clarity and consistency regarding what the term infographics, or information graphics, refers to in visualization research and practice. In particular, little is understood about where people's conceptualizations of infographics converge or diverge.
To address this gap, we conducted a systematic literature review and a practitioner survey to identify and contrast different perspectives on \igx.
We performed inductive coding on 487 sentences from 111 visualization and \hci papers, and analyzed questionnaire responses from 44 domain practitioners. Our findings reveal recurring dimensions 
{of conceptual disagreement on infographics: role of text, relationship with data visualization, and relationships with statistical charts and data comics.}}
{Based on the results, we recommend future work to explicitly report any assumptions made along these diverging conceptualizations, to investigate the cognitive origins of these disagreements, and to develop more holistic design component frameworks for infographics.}
{The \sm are available via {\osf} and as an \app.}
}


\begin{document}

\maketitle

\vspace{-2pt}
\section{Introduction}\label{sec:intro}

\Ig (or \igrx) are broadly used in research, education, and popular media. The term seems immediately understandable, yet often resists precise definition.
It is not difficult to find divergent or even contradictory statements on the origin and definition of \igx.
For example, Meggs~\etal wrote, \quote{The foundation for \textbf{information graphics} is analytic geometry, a branch of geometry developed \abbr by \abbr René Descartes. \abbr Cartesian coordinates and other aspects of analytic geometry were later used by \abbr William Playfair to convert statistical data into symbolic graphics. \abbr Playfair created a new category of graphic design, now called \textbf{information graphics}}~\cite{meggs2025meggs}. This view assumes that statistical graphics are inherently \igx. On the other hand, Chen~\etal considered \ig to be distinct from statistical charts. They classified narrative visualizations into six genres: annotated charts, \igx, timelines \& storylines, data comics, scrollytelling \& slideshow, and data videos, noting that \quote{existing tools for converting standard statistical charts into \textbf{infographics} support only simple chart conversions}~\cite{e006}. In this narrower definition, statistical data may be a necessary but not sufficient criterion for a design to qualify as \igx. 

\begin{figure}[!t]
    \centering
    \includegraphics[width=\columnwidth]{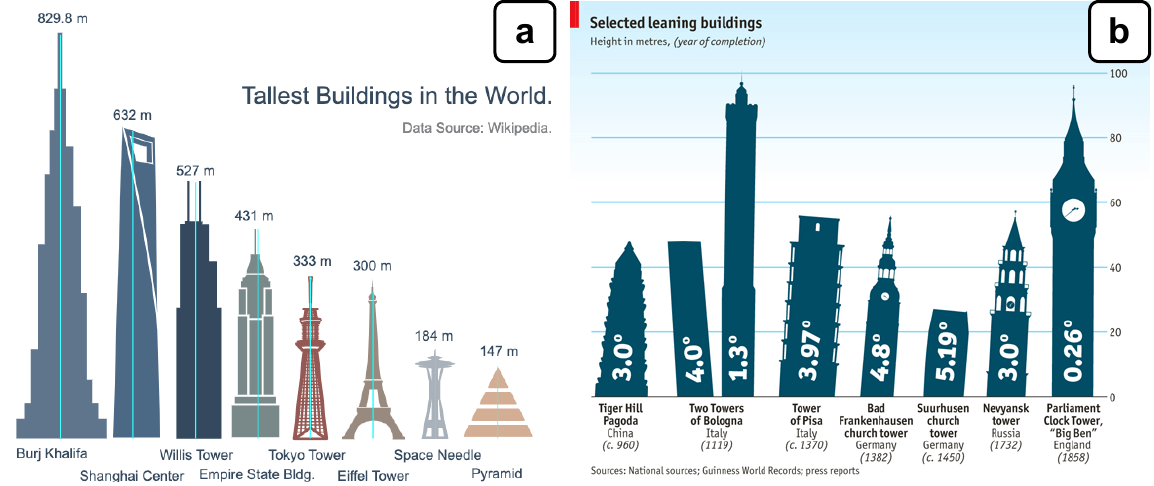}
    \vspace*{-18pt}
    \caption{(a)~\quote{Tallest Buildings in the World} by Kim \etalx~\cite{kim_data-driven_2016}. (b)~\quote{Selected leaning buildings} by The Economist online~\cite{economistTotteringTowers}. The two graphics are visually similar, {but (a) was considered \ig in~\cite{kim_data-driven_2016}, while (b) was not according to~\cite{stokes2025analysistextfunctionsinformation}.}}
    \label{fig-tower}
    \vspace*{-4pt}
\end{figure}

Some may attribute such inconsistencies to the differing disciplinary traditions and perspectives between graphic design and visualization.
However, even within the visualization community, researchers do not always agree on whether a specific design should be categorized as \igx. 
In an analysis of text functions in information visualization, Stokes~\etalx~\cite{stokes2025analysistextfunctionsinformation} decided to remove \ig from their corpus, but kept a graphic by the Economist (\cref{fig-tower}b), indicating that they did not consider the design to be an \igcx. In contrast, a similar design (\cref{fig-tower}a) was considered a prototypical example of \ig {by Kim~\etal in their work} on data-driven guides for designing expressive \igrx~\cite{kim_data-driven_2016}.
It is evident that when talking about \igx, individuals have different ideas and examples in mind.

\begin{figure}[!t]
    \centering
    \vspace*{-3pt}
    \includegraphics[width=\columnwidth]{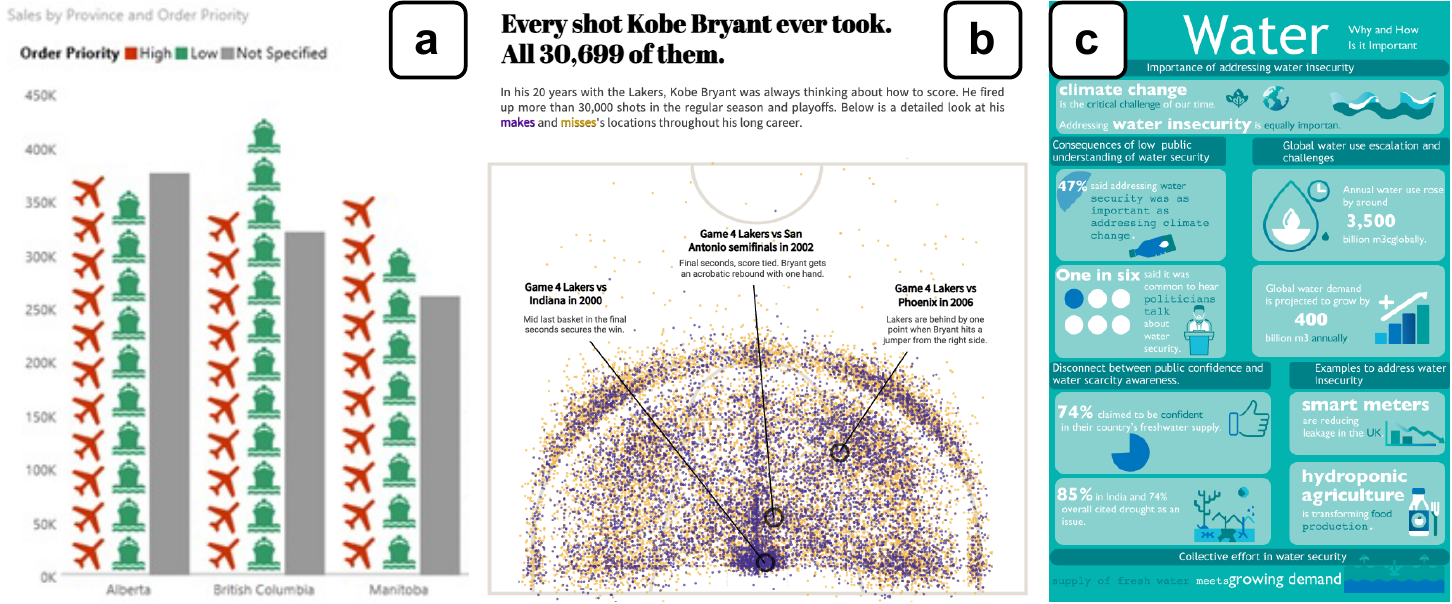}
    \vspace*{-18pt}
    \caption{{Example \ig produced by (a)~InfoNice~\cite{a058}, (b)~Epigraphics~\cite{a244} and (c)~InfoAlign~\cite{feng2026infoalign}. While all three authoring systems aim to support \igc creation, their design scopes vary significantly.}}
    \label{fig-meta}
    \vspace*{-16pt}
\end{figure}

{This conceptual open-endedness of \ig prevents cumulative scientific progress, because studies using the same term often investigate fundamentally different visual artifacts.
For example, several systems, including InfoNice~\cite{a058}, Epigraphics~\cite{a244}, and InfoAlign~\cite{feng2026infoalign}, have been developed to support \igc authoring. However, the example outputs presented in these papers reflect markedly different assumptions about what constitutes an \igc~(\cref{fig-meta}).
Consequently, it is difficult to compare or reason about the expressiveness of these systems, as they are designed for different target artifacts despite addressing the same stated problem.}

{While it has been recognized that no consensus on the definitive boundary of infographics exists~\cite{lankow2012infographics}, little prior work has systematically examined \emph{where} people's conceptualizations of \ig converge and diverge. Rather than prescribing a single definition, or simply observing the absence of consensus, we seek to systematically identify the dimensions along which these conceptualizations differ.}
Making these {dimensions} explicit is valuable not only for understanding the {conceptual} landscape, but also for revealing unresolved discrepancies and informing future research.
{We formulate our exploration as two research questions:}

\begin{itemize}[itemsep=-4pt, topsep=0pt, leftmargin=24pt]
\item[RQ1.] How do people conceptualize the term \flatquote{\igx}?
\item[RQ2.] How do people perceive \ig in relation to other visualization concepts?
\end{itemize}

We focus our investigation on two groups of stakeholders: (1)~visualization and \hci {\textbf{researchers}}, who drive research and applications related to \igx; and (2)~{\textbf{practitioners}} who use \ig in their own professional work.
For researchers, we conducted a systematic literature review, extracting and analyzing {their discussion of} \ig in 
{111} curated visualization\textor \hci papers.
For practitioners, we distributed a questionnaire via convenience sampling, gathering 44 responses that {(1)~define or describe \flatquote{\igx} and (2)~compare it with related visualization concepts}.
We present our findings on the shared and diverging understandings of \ig through inductive coding and {analysis of questionnaire responses}.
{Specifically, conflicting arguments arise along three major dimensions: (1)~the role of text in \igx, (2)~the relationship between \ig and data visualization, and (3)~whether statistical charts and data comics are considered \igx.}
{Based on these findings, we suggest that researchers should (1)~explicitly contextualize their work along these diverging conceptual dimensions, (2)~further investigate the cognitive sources of disagreements, and (3)~construct \igc design frameworks centered around the frequently discussed and debated components.}
\looseness=-10

\vspace{-2.5pt}
\section{Related Work}\label{sec:related-work}

A number of prior studies have performed meta-analyses on important visualization concepts to identify state-of-the-art research, synthesize findings, and highlight any disagreements or misconceptions about them.
{Concepts such as \emph{visualization novices}~\cite{y001}, \emph{visualization dashboards}~\cite{y002}, \emph{interaction in visualization}~\cite{y003}, \emph{multiscale visualization}~\cite{y004} and \emph{data storytelling}~\cite{y009} have been subjects of exploration through systematic literature reviews.}

However, {few} attempts have been made to systematically review the concept of \emph{\igx}.
{While many papers have investigated \igc application in pedagogical~\cite{y007a,y007b,y007c} or health contexts~\cite{y008} through literature surveys, their goal was mainly to synthesize a taxonomy of best practices for \igc usage, rather than to highlight diverging perceptions of the concept itself.
{In addition, we observe that these analyses were often based on \ig curated by inconsistent criteria.}
These criteria either originate from different sources} (\eg Jaleniauskiene~\etalx~\cite{y007b} uses definition from \cite{y007b-ref}, while Kong~\etalx~\cite{y008} quotes \cite{y008-ref-a,y008-ref-b})
{or involve a selective set of descriptive attributes that are essential to the study context (\eg \ig need to \quote{convey a message} and \quote{aid in data interpretation} in~\cite{y007c}, or to \quote{facilitate attractive information design} and \quote{enhance visual perception} in~\cite{y007a}).
Existing definitions of \ig also {trade specificity for generalizability:} Lankow \etal simply described them as \quote{\quotealt{using} visual cues to communicate information}~\cite{lankow2012infographics}; similarly, Krum characterized \ig as any graphic design with \quote{data visualizations, illustrations, text, and images \abbr that \quotealt{tell} a complete story}~\cite{krum2013infographics}.}
{These inconsistencies and ambiguities exemplify the gap in \igc perception within the visualization community.}
Our work bridges this gap by examining and summarizing the commonalities, ambiguities, and discrepancies in \ig understanding through an analysis of visualization and \hci literature and a practitioner questionnaire.

\vspace{-2.5pt}
\section{Methods}

In order to understand how researchers and practitioners conceptualize the term \flatquote{\igx,} we conducted two complementary studies: (1)~a systematic review of visualization/\hci literature and (2)~a practitioner questionnaire.

\vspace{-2pt}
\subsection{Literature Review}

\paragraph{Overview.}
We used the \textsc{prisma} 2020 statement for systematic reviews~\cite{prisma} to guide our paper selection process~(\cref{fig-prisma}).
We then extracted sentences from the screened papers that (1)~{define or describe the term \flatquote{\igx,}} (2)~provide examples of \igx, {\andorx} (3)~compare \ig to related concepts.
We elected for a more fine-grained sentence-level analysis~\cite{y001,y007b,y009} as opposed to the more common paper-level analysis~\cite{y003,y004,y007a,y007c,y008} in order to strengthen our focus on conceptual attributes of \ig regardless of the study context.
The sentence corpus was compiled for subsequent inductive coding aimed at identifying common themes and comparing and contrasting them.
\looseness=-1
\paragraph{Paper Collection.}
{Based on discussion between the two senior authors and preliminary keyword searches, we identified four sources} as a representative subset of visualization\textor \hci research on \ig and adjacent topics. {The sources} include ACM Digital Library~(%
major \hci venues like \textsc{chi}
and \textsc{uist}), %
selected IEEE venues~(%
\textsc{tvcg}, %
\textsc{cga}, %
and \textsc{iv}), %
CGF~({EuroVis, EuroGraphics}), %
and GI~({Graphics Interface Conference}).

{Through iterative discussion between the first author and two senior authors,} we identified the following 9 keywords: \quote{\igcx(s),} \quote{\igrx,} \quote{visual narrative,} \quote{narrative visualization,} \quote{data video,} \quote{data comic(s),} \quote{glyph,} \quote{storytelling,} and \quote{graphic design.}
{These terms are often used interchangeably or together with \ig in literature.}
We performed a keyword search in all four sources to find papers that include any of the 9 keywords in their title or abstract. In total, we collected 807 papers published between 1971 and 2025. Full text and metadata of all papers were downloaded for further analysis.
\looseness=-10

\begingroup
\setlength{\columnsep}{8pt}
\begin{wrapfigure}[29]{r}{120pt}
    \centering
    \vspace{-11pt}
    \includegraphics[width=115pt]{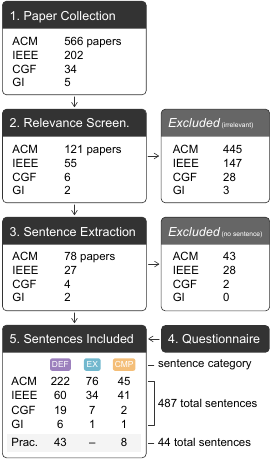}
    \vspace{-8pt}
    \caption{\textbf{Our method overview.} We collected sentences about \ig from a three-stage literature review and responses {from a practitioner (Prac.)~questionnaire. One sentence may be assigned multiple categories, so the total is smaller than the sum of individual counts.}}
    \label{fig-prisma}
\end{wrapfigure}
\aftergroup
\leavemode

\paragraph{Relevance Screening.~}
All seven authors participated in paper screening to determine their relevance by consulting the full text. {They were randomly assigned with papers for equal work division.} Due to the substantial size of the literature pool, each paper was assigned to exactly one author.
Papers whose research did not focus on \ig or include \ig as a subject of study (despite mentioning one or more related concepts in their title or abstract) were discarded from the corpus.
After the screening, 184 papers (22.8\% of the initial total) remained in the corpus.

\paragraph{Sentence Extraction.~}
The five student authors extracted sentences from the remaining papers; they were randomly assigned to papers for equal work division.
We targeted sentences that either (1)~{give a \emph{definition}~(\typea) or describe certain characteristics of \igx;} (2)~provide {or} describe one or more \emph{examples}~(\typeb) of \igx; or (3)~{\emph{compare}~(\typec) and contrast \ig against another visualization concept.} Each extracted sentence should fall into at least one of these three categories.
If one paper contained more than 10 sentences that satisfied the above criteria, only the 10 most distinct instances {(selected based on content diversity and section coverage)} were included.
If the paper did not have any sentence that satisfied the above criteria, it was discarded from the corpus.
In total, we extracted 487 sentences from 111 papers (60.3\% of the previous total). Among these sentences, 307 were categorized as \typea, 118 as \typeb\ and 89 as \typec.\looseness=-10

\vspace{-2pt}
\subsection{Practitioner Questionnaire}
To complement our findings from the literature review, {we designed and distributed an online questionnaire to gather first-hand insight from domain practitioners on their perceived definition and boundary of \igx. This study was approved by the university IRB.}\looseness=-10

\paragraph{Questionnaire Design.}
{In the questionnaire, we ask about the participant's occupation and domain, followed by two \igcx-related questions.}
Q1 requires participants to provide a short textual description or definition for \ig in their own words. Q2 focuses on participants' perception of the relationship between \ig and related concepts.
From the 487 sentences extracted from the literature review, we identified 27 commonly discussed {visualization concepts} through term frequency analysis (\eg \flatquote{chart,} \flatquote{data comic,} \flatquote{visualization;} see \cref{fig-concepts}). Excluding \flatquote{\igx} and its derivatives, the remaining 26 concepts were used as candidates to be compared against \ig in Q2.
Each participant received 10 randomly sampled concepts, ensuring that responses were evenly distributed. {They would determine the set relation between \ig and each given concept:} \emph{superset}, \emph{subset}, \emph{equivalent}, \emph{intersecting}, \emph{disjoint}, or \emph{unsure}.
\looseness=-10

\paragraph{Participant Recruitment.}
Participants were required to (1)~be at least 18 years of age and (2)~either consume, create, or use \ig as part of their professional work. We distributed the questionnaire via (1)~departmental graduate student and faculty mailing lists at colleges of computer science, social science and business at the authors' institution; (2)~public mailing lists and Slack channels for \hci and visualization communities; and (3)~the authors' personal networks.
The participation was voluntary with no financial compensation. {Participants provided consent by proceeding with the questionnaire.} 
In total, we received 44 responses, yielding 44 \igc definitions (43 categorized as \typea\ and 8 as \typec) and 440 concept relation votes (avg.~16.9 votes per concept).
{Surveyed occupations include student, research faculty, scientist, designer, educator, journalist and engineer, primarily from science, business and education domains.}

\vspace{-2pt}
\subsection{{Inductive Coding}}
We combined 487 {extracted sentences} from the literature review and 44 {definitions or descriptions} from the practitioner questionnaire into a corpus of 531 sentences. All five student authors participated in a 2-round inductive coding on this corpus. Sentences were randomly assigned to coders for equal work division. Two coders started by coding each sentence independently; in case of discrepancies, a third coder would step in to reach a consensus.
We developed a codebook for each of the three sentence categories, which are not mutually exclusive.
We also identified major \emph{contrary} relationships between codes, where pairs or groups of codes would make contradicting arguments.
{Overall,} this process yielded 102 codes (46 from \typea\ sentences, 24 from \typeb, and 32 from \typec)\footnote{{To facilitate exploration of our findings, we further categorized these codes into 14 high-level themes and 6 top-level aspects. See our web application and \sm for details.}} and 12 sets of \emph{contrary} relationships.
{The first two coders achieved an average inter-rater reliability (we report Gwet's AC1 due to the imbalanced distribution of code occurrences~\cite{Quarfoot01102016}) of 0.939 over all codes (0.937 for \typea, 0.953 for \typeb, 0.932 for \typec).}

\vspace{-2pt}
\section{Findings}

\subsection{Frequently Appearing Codes}
We present the 5 most frequently discussed code topics, each with over 40 total occurrences across \typea, \typeb\ and \typec\ sentences.
{The number of unique papers (\countl) and participants (\countp) for each code are also reported.}
These codes reflect some of the most prominent characteristics commonly associated with \ig {in literature.}

\code{{\anotherpoint}\Ig is a type of data visualization \textnormal{(\codefrom{\nc{21}}{13}{3})} that visualizes data or the relationship between data} %
{\codefrom{\na{77}}{41}{20}}
{\group{Design}{Definition}, \group{Comparison}{Relationship with other concepts}}
{Surveyed papers generally agree that \ig is a subtype of data visualization, {which translates data into graphical representations that facilitate understanding and reveal underlying connections between them~\cite{a054}.}}

\code{{\anotherpoint}\Ig contain diverse and effective visual representations or data encodings} %
{\codefrom{\na{45}, \nb{20}}{37}{15}}
{\group{Design}{Visual elements\textor components}}
{{\Ig are composed of a rich set of design components that facilitate reader's comprehension, engagement and memorability~\cite{u006}. Some notable features include pictographic marks~\cite{a058}, artistic decorations~\cite{e115}, and explanatory annotations (P3,~\cite{a058}).}}

\code{{\anotherpoint}\Ig can facilitate and accelerate data communication} %
{\codefrom{\na{91}}{44}{23}}
{\group{Goal}{Effects on readers}}
{{\Ig are frequently described as a more efficient means of data communication than statistical charts, since they present data and knowledge in a concise and straightforward manner~\cite{a422} that allows them to be more easily discovered and digested by the general public~\cite{a286}.}}

\code{{\anotherpoint}\Ig have a certain high-level narrative goal} %
{\codefrom{\na{55}}{29}{11}}
{\group{Goal}{Narrative goals}}
{{The goal of \ig is often to tell a story or deliver a message supported by statistics to the intended audience~\cite{a054}. This high-level narrative may dictate the overall theme and appearance of an \igc in its design phase~\cite{a308}.}}

\code{{\anotherpoint}\Ig are broadly used for education \textnormal{(\codefrom{\na{36}}{16}{1})} and in popular media} %
{\codefrom{\na{25}}{21}{1}}
{\group{Application}{Application domains}, \group{Application}{Practical use cases}}
{{%
\Ig are commonly produced for education purposes and distributed via popular media (\eg newspapers and magazines)~\cite{e006}. They help students and the general public acquire new information through a blend of text and visual content~\cite{a536}.}}

\vspace{-2pt}
\subsection{Contrary Relations}\label{sec-code-relation}
We also present the three most prominent \emph{contrary} relations between sets of codes. These codes highlight emerging debates on the prototypical design patterns of \ig and different perspectives about its conceptual boundary.
\paragraph{{\anotherpoint}Role of text elements in \igx.}
Some works and participants propose that \codetext{both text and graphical elements are important for \igx} (\codefrom{\na{36}, \nb{5}, \nc{11}}{35}{7}), defining \ig as \quote{a popular medium for storytelling with visual elements around text messages}~\cite{e042} and which \quote{contains a mixture of charts, text, and graphics}~(P41). {A combination of both text and visuals is believed to engage readers and promote their overall understanding of the work~\cite{e053}.
Another opposing argument is} that \codetext{\ig {should} prioritize visual content over text in their designs} (\codefrom{\na{6}, \nb{2}, \nc{3}}{8}{1}), as exemplified by statements like \quote{\abbr all effective infographics \abbr contain minimal text to display key points supported by corresponding graphics}~\cite{a422} and \quote{\quotealt{infographics} are a way to tell a story or explain a concept using images, charts, and minimal text}~(P7). They emphasize that the fundamental role of \ig is to {show key insight} efficiently through visual representations, {whereas texts often require more mental effort in order to digest~\cite{e005}.}
However, {we noticed that the \igc quoted by Bellato~\cite{a422} as having \quote{minimal text}~(\cref{fig-text}a) may actually contain} more text than what Holloway~\etalx~\cite{a298} considered a standard news \igc with \quote{a mix of text and graphics}~(\cref{fig-text}b). This signifies that the perceptual standard for text density in \ig may not be always consistent across works.

\begin{figure}[!tbp]
    \centering
    \includegraphics[width=\columnwidth]{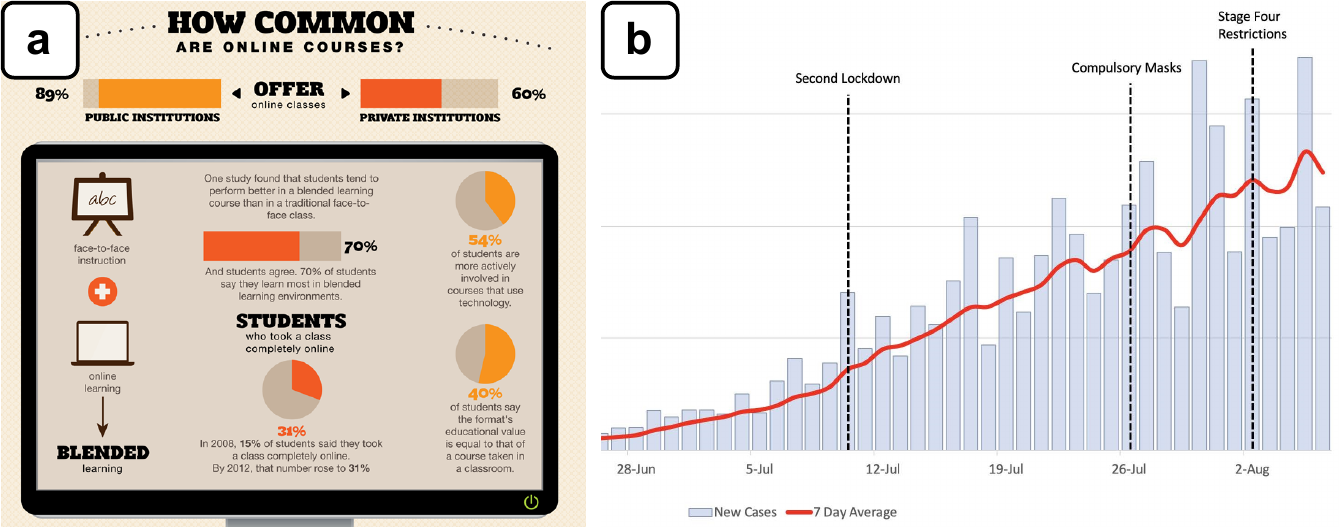}
    \vspace*{-18pt}
    \caption{(a)~\quote{Why Online Learning Is Vital to Improving Education} by Online Degree Programs~\cite{OnlineDegreeProgramsIgx}. (b)~A news \igc on COVID-19 number of cases over time. While (a) was described as having \quote{minimal text} according to~\cite{a422}, {(b) features even less text by word count and density but was still considered \quote{a mix of text and graphics} in~\cite{a298}. Both images are cropped due to long aspect ratios.}\looseness=-10}
    \label{fig-text}
    \vspace*{-16pt}
\end{figure}

\paragraph{{\anotherpoint}Relationship between infographic and data visualization.}
{Many papers and participants believe} that \codetext{\ig are a kind of data visualization} (\codefrom{\nc{21}}{13}{3}). {This is evident from statements} such as \quote{Included in this discussion of data visualization is a subset of data visualization \--- infographics}~\cite{a457},
\quote{Data visualization includes broad categories of spatial representation such as \abbr infographics}~\cite{a557},
{and} \quote{infographics are a combination of data visualization and graphic design}~\cite{e027}.
Yet, {others} have conversely stated that \codetext{\ig are different from data visualization} (\codefrom{\nc{8}}{5}{1}). P12 {argues} that \quote{Informative graphics \quotealt{are} not necessarily data visualization.} Holloway~\etal echoed {their view}, pointing out that traditional data visualizations \quote{do not come with a message: it is up to the viewer to analyse and explore the data and come to their own conclusions}~\cite{a298}. \Igx, on the other hand, are {narrative-driven, characterized by an intended message} which the author aims to get across to the {target audience}.

\paragraph{{\anotherpoint}Relationship between \igx, statistical charts and data comics.}
One {opinion} is that \codetext{\ig include statistical charts} (\codefrom{\na{29}, \nc{20}}{30}{12}). A {large number} of papers adopt the definition that \ig are \quote{non-pictorial graphics such as bar charts and line graphs}~\cite{a054}, emphasizing that both {\ig and statistical charts} avoid using visual elements that resemble real-life objects (\quote{non-pictorial}) as part of their design.
{Some others suggest a stricter criteria:} \codetext{infographics should only include embellished statistical charts} (\codefrom{\na{21}, \nc{9}}{20}{3}). {Qualifying designs should superimpose useful embellishments (\eg annotations and decorations) over basic charts that} \quote{help readers easily interpret the story}~\cite{a340}, \quote{convey abstract information appealingly}~\cite{e115}, and \quote{convey data stories in a more compelling and narrative-driven way}~(P20).
Another view is that \codetext{\ig include data comics} (\codefroml{\nc{1}}{1}), {proposed exclusively as part of the Epigraphics taxonomy~\cite{a244}}. The author identified data comics as a form of \flatquote{directive \igx,} which utilizes sequential ordering of visual components to represent logic flow.
{Finally, one view} refutes all of the above: \codetext{\ig are a different type of data visualization from statistical charts and data comics} (\codefroml{\nc{19}}{10}). {While an explicit claim is rare,} many studies have considered \igx, statistical charts and data comics as distinct visualization categories when building taxonomies~\cite{a136}, conducting surveys~\cite{a557,e006} or developing authoring schemes~\cite{e007}.
\vspace{-2pt}
\subsection{Concept Relations}\label{sec-term}
{From the practitioner questionnaires, we obtained the perceived set relations between \ig and 26 related concepts across 44 participants. \cref{fig-concepts} shows the distribution of participant votes. We discuss some notable patterns below.}

{First, while the perceived relation for \flatquote{(data, information) visualization} is fuzzy, a majority of participants agreed that \flatquote{\igx} is related to \flatquote{(data) visualization} and is equivalent to \flatquote{information visualization.}
{Second, all 6 concepts describing generic or specific types of statistical charts (\eg bar chart, scatter plot) were considered subsets of \flatquote{\igx} as the majority relation.}
Third, participants were unable to reach a consensus for \flatquote{cartoon,} \flatquote{pictograph} and \flatquote{figure.} Noticeably, \flatquote{pictograph} is the only concept with \emph{unsure} being one of its majority relations, suggesting participants' uncertainty or unfamiliarity with this concept despite its appearances in \igc literature.
Finally, \emph{intersecting} is the only relation that received non-zero votes for all 26 concepts, at a minimum of 15.8\% votes. This implies that both our reviewed literature and {surveyed} practitioners identify these concepts as being relevant to \igx.}\looseness=-10

\begin{figure}[!tbp]
    \centering
    \vspace*{-8pt}
    \includegraphics[width=\columnwidth]{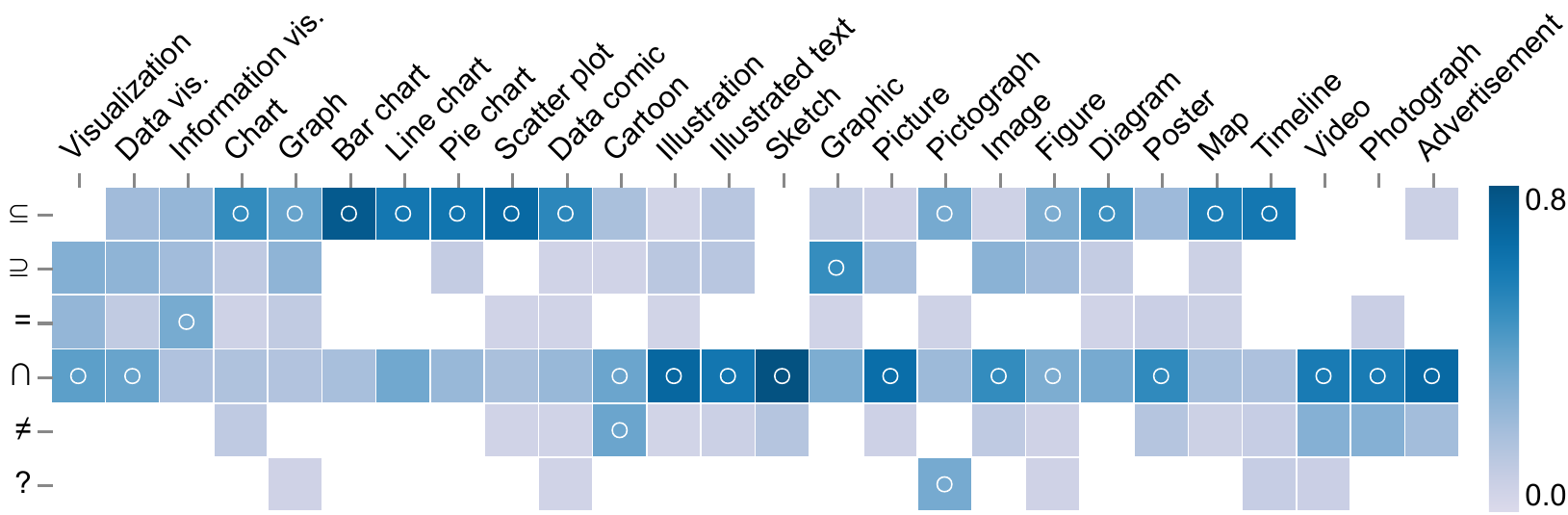}
    \vspace*{-18pt}
    \caption{{Frequency of set relation votes between \flatquote{\igx} ($I$) and 26 related concepts ($X$). The 6 set relations are \emph{superset} ($X\subseteq\nolinebreak I$), \emph{subset} ($X\supseteq\nolinebreak I$), \emph{equivalent} ($X=\nolinebreak I$), \emph{intersecting} ($X\cap\nolinebreak I$), \emph{disjoint} ($X\ne\nolinebreak I$) and \emph{unsure} ($?$). Blank cells indicate zero votes, and a circle ($\circ$) indicates majority vote for a given concept. Values of each column add up to 1.}}
    \looseness=-10
    \label{fig-concepts}
    \vspace*{-16pt}
\end{figure}

\vspace{-2pt}
\section{{Implications for Research and Practice}}

\paragraph{{Clarifying Assumptions for \Igc Investigation.}}
{Some prior work has recognized that a universally accepted definition for \ig is unlikely to exist; for example, Lankow~\etal remarked that \quote{There is no threshold at which something `becomes' an infographic}~\cite{lankow2012infographics}.
Our findings build on such observations by identifying recurring dimensions along which conceptual disagreements arise, \eg the role of text {and relationships with data visualization, statistical charts and data comics.}
Different positions along these dimensions influence a wide range of research activities, including corpus curation, system design, and empirical validation. We thus suggest that researchers should explicitly contextualize their work along these {diverging conceptual dimensions}.
For example, papers should specify how \ig are positioned relative to adjacent visualization concepts, whether basic or embellished statistical charts are considered during sample curation, and what role textual, narrative, and illustrative elements play in the characterization of representative designs. By making these assumptions explicit, readers will be able to better distinguish between differences that arise from {conscious methodological choices of a certain research} versus those generated by fundamentally different conceptualizations of \igx.}

\paragraph{Understanding Cognitive Sources of Disagreement.}
{Our findings suggest that infographics categorization may be guided by  examples rather than explicit definitions. For example, in \cref{fig-concepts}, the participants reached stronger consensus for the relation of concrete chart types such as \flatquote{bar chart} and \flatquote{scatter plot} than for \flatquote{chart} and \flatquote{graph} in general. This observation is consistent with prototype theory~\cite{rosch1975cognitive}, which posits that humans categorize objects based on {\emph{prototypes}} \--- real-world examples that best represent a given category.
We hypothesize that the diverging conceptualizations on \ig arise due to the different prototypical examples that individuals associate with this term.
Future work should investigate these prototypes through interviews and focus groups to better understand the cognitive origins of disagreement.}

\paragraph{Toward Holistic Component Frameworks for \Igx.}
Our findings reveal that people characterize \ig through multiple interacting aspects rather than standalone features
{(\eg a mix of text density, embellishments, non-pictorial elements, logic flow, and layout).}
{However, existing taxonomies for \ig and related concepts mostly only focus on one single dimension of design (\eg data mark~\cite{a058}, path layout~\cite{a340}, or metaphor~\cite{e060}) or certain functional aspects (\eg purpose and audience~\cite{y002}, or presentation modality~\cite{y003,y009}).}
{In this work, the frequent and conflicting codes we discovered highlight a range of important design components (\eg texts, pictograms, and annotations) that sway people's perception of \igx.
A comprehensive component framework that incorporates these observations} could provide a more precise vocabulary for describing \igc designs, support more systematic comparison of authoring systems and datasets, and establish a stronger conceptual foundation for \igc research.
\looseness=-10

\vspace{-2pt}
\section{Conclusion}

{Through a systematic analysis of literature and practitioner perspectives, we characterized the conceptual landscape of \igx. We identified recurring dimensions that converge or distinguish between different conceptualizations.}
{We hope that these findings will encourage more explicit communication of methodological assumptions, inspire research into the cognitive origins of disagreements, and support the development of richer, component-based frameworks for \igx.}

\acknowledgments{{We would like to thank all our lab members and the anonymous reviewers for their constructive feedback on this work.
We would like to thank all questionnaire participants for their valuable input. This work was supported by NSF grant IIS-2239130.}}

\enlargethispage{\baselineskip}

\bibliographystyle{abbrv-doi}
\bibliography{ref}

\end{document}